\documentclass[letterpaper]{article}
\usepackage{alife9}
\usepackage{color}

\title{Evolution of Robust Developmental Neural Networks}
\author{Alan N. Hampton$^{1}$ \and Christoph Adami$^{1,2}$ \\
\mbox{} \\
$^1$Digital Life Laboratory 136-93, California Institute of Technology, Pasadena, CA 91125\\
$^2$Jet Propulsion Laboratory 126-347, California Institute of
Technology, Pasadena, CA 91109\\
adami@caltech.edu}

\begin{document}
\maketitle

\begin{abstract}
We present the first evolved solutions to a computational task
within the {\emph{N}}euronal {\emph{Org}}anism {\emph{Ev}}olution
model (\textbf{\it Norgev}) of artificial neural network
development. These networks display a remarkable robustness to
external noise sources, and can regrow to functionality when
severely damaged.  In this framework, we evolved a doubling of
network functionality (double-NAND circuit).  The network
structure of these evolved solutions does not follow the logic of
human coding, and instead more resembles the decentralized
dendritic connection pattern of more biological networks such as
the \mbox{{\it C. elegans}} brain.
\end{abstract}

\section{Introduction}

The complexity of mammalian brains, and the animal behaviors they
elicit, continue to amaze and baffle us.  Through neurobiology, we
have an almost complete understanding of how a single neuron
works, to the point that simulations of a few connected neurons
can be carried out with high precision. However, human designed
neural networks have not fulfilled the promise of emulating these
animal behaviors.

The problem of designing the neural network {\em structure} can be
generalized to the problem of designing complex computer programs
because, in a sense, an artificial neural network is just a
representation of an underlying computer program. Computer
scientists have made substantial progress in this area, and
routinely create increasingly complicated codes. However, it is a
common experience that when these programs are confronted with
unexpected situations or data, they stall and literally stop in
their tracks. This is quite different from what happens in
biological systems, where adequate reactions occur even in the
rarest and most uncommon circumstances, as well as in noisy and
incompletely known environments. It is for this property that some
researchers have embraced evolution as a tool for arriving at
robust computational systems.

Darwinian evolution not only created systems that can withstand
small changes in their external conditions and survive, but has
also enforced {\em functional modularity} to enhance a species'
evolvability \cite{Gerhart98} and long-term survival. This
modularity is one of the key features that is responsible for the
evolved system's robustness: one part may fail, but the rest will
continue to work. Functional modularity is also associated with
component re-use and developmental evolution \cite{Koza03}.

The idea of evolving neural networks is not new
\cite{Kitano90,Koza91}, but has often been limited to just
adapting the network's structure and weights with a bias to
specific models (e.g., feed-forward) and using {\em homogeneous}
neuron functions. Less constrained models have been proposed
\cite{Belew93,Eggenberger97,Gruau95,Nolfi95}, most of which
encompass some sort of implicit genomic encoding. In particular,
developmental systems built on artificial chemistries (reviewed in
Dittrich et al. 2001)\nocite{Dittrich01} represent the least
constrained models for structural and functional growth, and thus
offer the possibility of creating modular complex structures.
Astor and Adami (2000) \nocite{Astor00} introduced the
\textbf{Norgev} (\textbf{\emph{N}}euronal \textbf{\emph{Org}}anism
\textbf{\emph{Ev}}olution) model, which not only allows for the
evolution of the developmental mechanism responsible for the {\em
growth} of the neural tissue or artificial brain, but also has no
\emph{a priori} model for how the neuron computes or learns. This
allows neural systems to be created that have the potential of
evolving developmental robustness as found in nature.  In this
paper, we present evolved neural networks using the Norgev model,
with inherent robustness and self-repair capabilities.

\section{Description of Norgev}

Norgev is, at heart, a simulation of an artificial wet chemistry
capable of complex computation and gene regulation. The model
defines the tissue substrate as a two-dimensional hexagonal grid
on which proteins can diffuse through discrete stepped diffusion
equations. On these hexagons, neural cells can exist, and carry
out actions such as the production of proteins, the creation of
new cells, the growth of axons, etc. Proteins produced by the cell
can be external (diffusible), internal (confined within the cell
and undiffusible) or neurotransmitters (which are injected through
connected axons when the neuron is excited). Cells also produce a
constant rate of cell-tag proteins, which identify them to other
cells and diffuse across the substrate.

\begin{figure}[h]
\begin{center}
\includegraphics[height=3.1in,angle=-90]{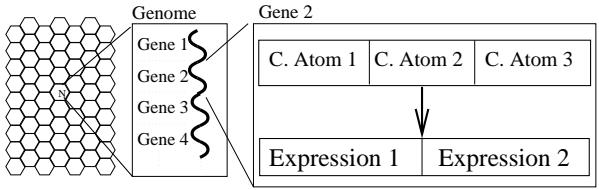}
\caption{Neural cells are placed into an hexagonal grid and then
start executing their genome, which consists of a series of
conditions followed by a series of expression actions.}
\label{DNAexec}
\end{center}
\end{figure}

Each neural cell carries a genome which encodes its behavior.
Genomes consist of genes which can be viewed as a genetic program
that can either be executed (expressed) or not, depending on a
gene {\em condition} (see Fig.~\ref{DNAexec}).  A gene condition
is a combination of several condition {\em atoms}, whose values in
turn depend on local concentrations of proteins. The gene
condition can be viewed as the upstream regulatory region of the
genetic program it is attached to, while the atoms can be seen as
different binding modules within the regulatory region. Each gene
is initially active (activation level $\theta=1$) and then each
condition atom acts one after another on $\theta$, modifying it in
the $[0,1]$ range, or totally suppressing it ($\theta=0$). Table
\ref{cond1} shows all the possible condition atoms and how they
act on the gene expression level $\theta$ passed on to them.

\begin{table}[h]
\center{ \small
\begin{tabular}{|l|l|}
  \hline
  Cond. Atom & Evaluation value $\theta$\\
  \hline
  SUP[\emph{CTPx}] & $=\theta$, if cell \textbf{is} of type \emph{CTPx} \\
  NSUP[\emph{CTPx}] & $=\theta$, if cell \textbf{is not} of type \emph{CTPx} \\
  ANY[\emph{PTx}] & $=\theta$, if [\emph{PTx}]$\neq 0$ \\
  NNY[\emph{PTx}] & $=\theta$, if [\emph{PTx}]$=0$ \\
  \hline
  ADD[\emph{PTx}] & $=R^1_0(\theta + $[\emph{PTx}]$)$ \\
  SUB[\emph{PTx}] & $=R^1_0(\theta - $[\emph{PTx}]$)$ \\
  MUL[\emph{PTx}] & $=\theta*$[\emph{PTx}] \\
  AND[\emph{PTx}] & $=min(\theta,$[\emph{PTx}]$)$ \\
  OR[\emph{PTx}] & $=max(\theta,$[\emph{PTx}]$)$ \\
  NAND[\emph{PTx}] & $=1-$AND[\emph{PTx}] \\
  NOR[\emph{PTx}] & $=1-$OR[\emph{PTx}] \\
  NOC[\emph{PTx}] & $=\theta$, the neutral condition \\
  \hline
\end{tabular}
} \caption{Repressive and evaluative condition atoms:  The SUP and
NSUP conditions evaluate the cell-type of the cell in which they
are being executed.  On the other hand, ANY and NNY repress the
gene under the influence of {\em any} type of protein (internal,
external, cell-type or neurotransmitter), where '[\emph{PTx}]'
stands for the concentration of protein \emph{PTx}. The neutral
condition is special and acts as a silent place holder. $R^1_0()$
saturates the activation into the [0,1] range.} \label{cond1}
\end{table}

Once a gene activation value $\theta$ has been reached, each of
the gene's expression atoms are executed.  Expression atoms can
carry out simple actions such as producing a specific protein, or
they can emulate complex actions such as cell division and axon
growth.  Table \ref{expr1} contains a complete list of expression
atoms used in Norgev. A more complete description of the Norgev
model and its evolution operators (mutation and crossover) can be
found in \cite{Astor00}.

\begin{table}[h]
\center{ \small
\begin{tabular}{|l|l|}
  \hline
  Expr. Atom & Action description\\
  \hline
  PRD[\emph{XY}] & produces substrate \emph{XY}\\
  SPL[\emph{CTPx}] & divide. offspring of type \emph{CTPx} \\
  GRA[\emph{XY}] & grow axon following \emph{XY} gradient \\
  GDR[\emph{XY}] & grow dendrite following \emph{XY} gradient  \\
  EXT & excitory stimulus \emph{XY} \\
  INH & inhibitory stimulus \emph{XY} \\
  MOD+[\emph{NTx}] & increase connection weights \\
  MOD-[\emph{NTx}] & decrease connection weights \\
  RLX[\emph{NTx}] & relax weights \\
  DFN[\emph{NTx}] & define cells neurotransmitter \\
  NOP & null action, neutrality \\
  \hline
\end{tabular}
}  \caption{Expression atoms.  Each is influenced by $\theta$ in a
different way. For PRD it states the production quantity; for SPL,
GRA and GDR the probability of execution; for EXT and INH the
stimulus amount; for MOD+, MOD- and RLX the increase, decrease and
multiply factor; and for DFN and NOP, $\theta$ has no influence.}
\label{expr1}
\end{table}

We know that in cellular biology, gene activation leads to the
production of a specific protein that subsequently has a function
of its own, ranging from enzymatic catalysis to the docking at
other gene regulatory sites. In this model, the most basic
expression element is the production of proteins (local or
externally diffusible) through the PRD[\emph{PTx}] atom.  These
can then interact and modulate the activation of other genes in
the genome. In this sense, it can be argued that they are only
regulatory proteins. However, at least abstractly, genes in this
model need not only represent genes in biological cells but can
also represent the logic behind enzyme interaction and their
products. Thus, Norgev's genome encodes a dynamical system that
represents low level biological DNA processes, as well as higher
level enzymatic processes including long-range interaction through
diffusible substances like hormones.  However, the objective is
not to create a complete simulation of an artificial biochemistry,
and thus other expression atoms are defined that represent more
complex actions, actions that in real cells would need a whole
battery of orchestrated protein interactions to be accomplished.

\section{Organism example}

The best way to understand the model is probably to sit down and
create by hand a functional organism.  Here we will present a
handwritten organism (Fig.~\ref{sgenome}) and explain how it
develops into a fully connected neural network that computes a
NAND logical function on its two inputs and sends the result to
its output.

\begin{figure}[h]
\begin{center}
\colorbox[gray]{0.9}{\scriptsize \setlength{\tabcolsep}{0.2mm}
\begin{tabular}{llll}
  1. & SUP(\emph{cpt}) ANY(\emph{cpt})  & $\Rightarrow$ & SPL(\emph{acpt0}) \\
  2. & SUP(\emph{acpt0}) ADD(\emph{apt0}) SUB(\emph{cpt})  & $\Rightarrow$ & SPL(\emph{acpt3})\\
  3. & SUP(\emph{acpt0}) ADD(\emph{spt0}) SUB(\emph{spt1})  & $\Rightarrow$ & SPL(\emph{acpt1}) \\
  4. & SUP(\emph{acpt0}) ADD(\emph{spt1}) SUB(\emph{spt0})  & $\Rightarrow$ & SPL(\emph{acpt2}) \\
  5. & SUP(\emph{acpt0}) ADD(\emph{cpt}) & $\Rightarrow$ & SPL(\emph{acpt0}) \\
  6. & SUP(\emph{acpt1}) ANY(\emph{spt0})  & $\Rightarrow$ & GDR(\emph{spt0})  DFN(\emph{NT1}) \\
  7. & SUP(\emph{acpt2}) ANY(\emph{spt1})  & $\Rightarrow$ & GDR(\emph{spt1})  DFN(\emph{NT2}) \\
  8. & SUP(\emph{acpt3}) ANY(\emph{apt0})  & $\Rightarrow$ & GDR(\emph{acpt1})  GDR(\emph                                                                   {acpt2})  GRA(\emph{apt0}) \\
  9. & SUP(\emph{acpt3}) ADD(\emph{NT1}) NAND(\emph{NT2})  & $\Rightarrow$ & EXT0 \\
  10. & ANY(\emph{eNT})  & $\Rightarrow$ & EXT0 \\
\end{tabular}
} \caption{Genome of \emph{\textbf{Stochastic}}} \label{sgenome}
\end{center}
\end{figure}

\begin{figure}[h]
\begin{center} \setlength{\tabcolsep}{0.2mm}
\begin{tabular}{cc}
time = $4$ & time = $24$ \\
\includegraphics[width=1.65in]{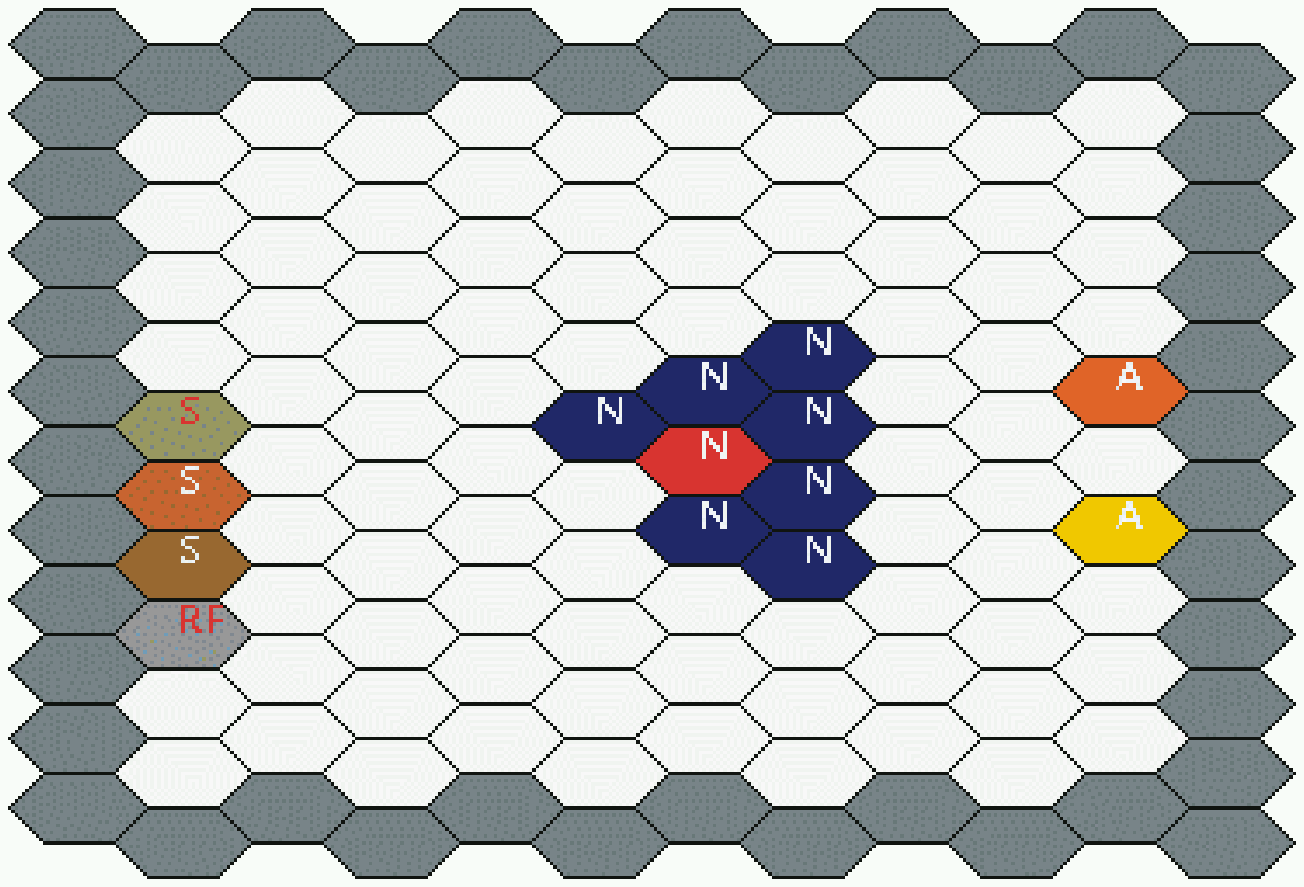} &
\includegraphics[width=1.65in]{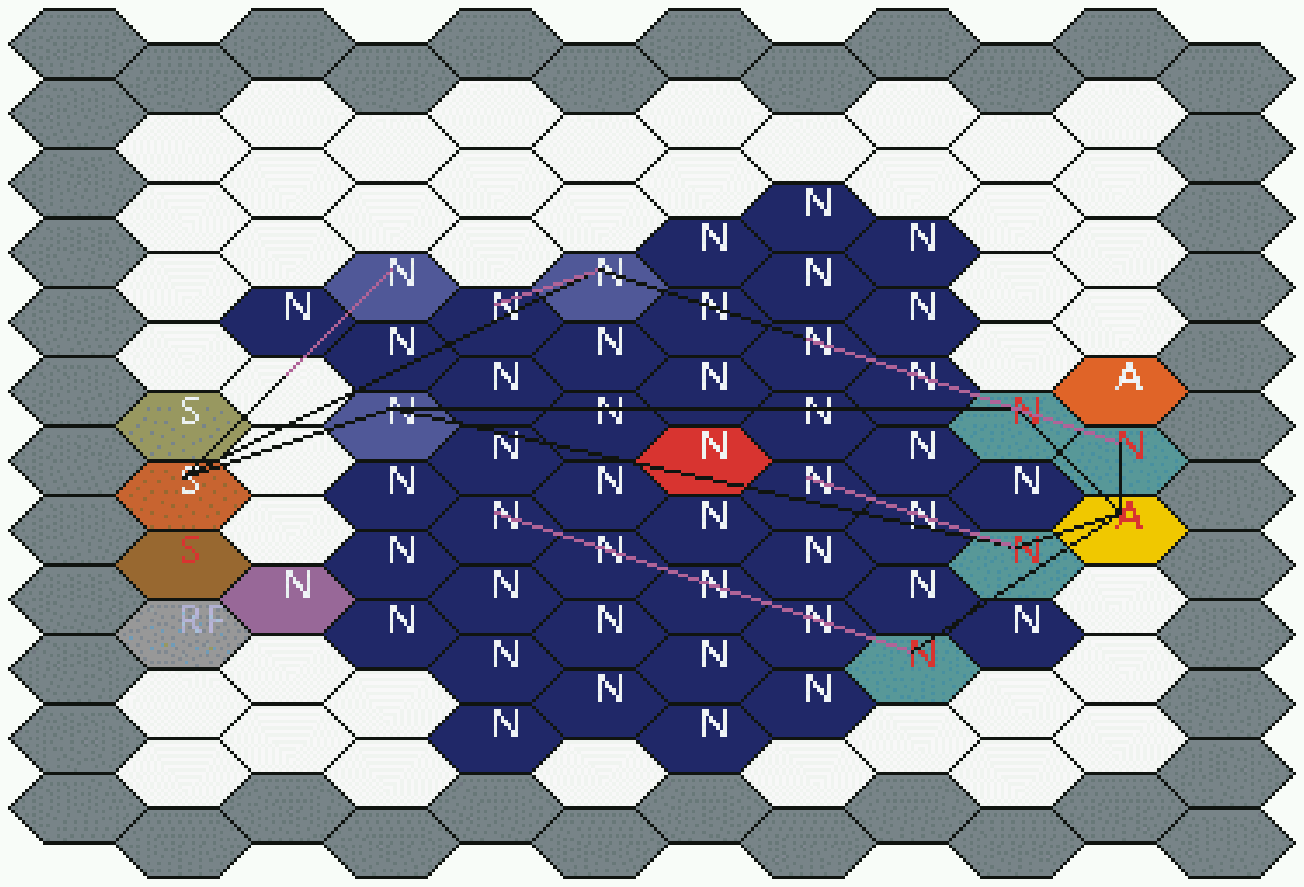} \\
 time = $40$ & time = $120$ \\
\includegraphics[width=1.65in]{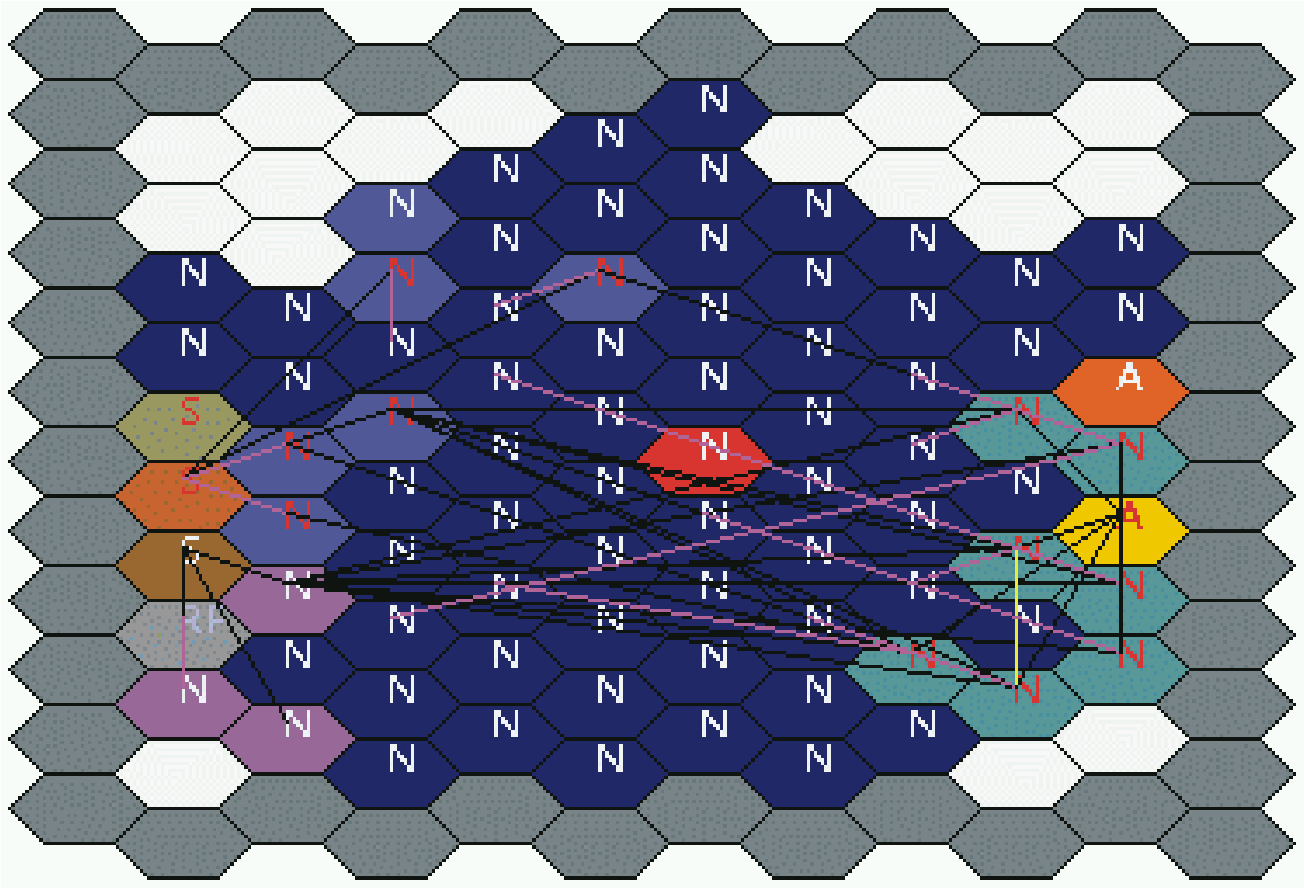} &
\includegraphics[width=1.65in]{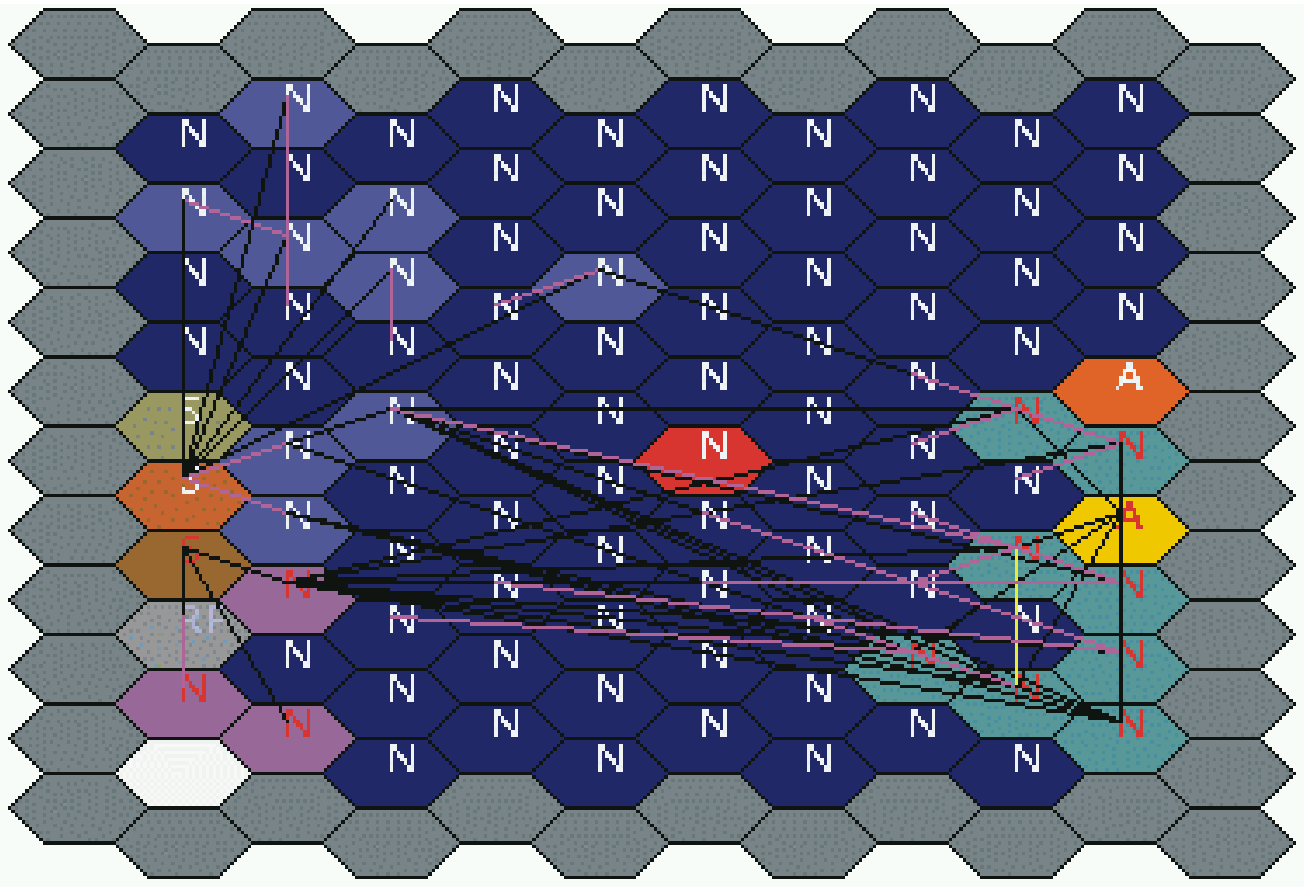} \\
\end{tabular}
\caption{Successive stages in the developmental growth of the
\emph{\textbf{Stochastic}} neural tissue.} \label{cellSto}
\end{center}
\end{figure}

The organism, which we named \emph{\textbf{Stochastic}}, relies on
the random nature of the underlying chemical world to form its
tissue structure. When an organism is first created, a tissue seed
(type \emph{CPT}) is placed in the center of the hexagonal grid,
two sensor cells on the left of the grid and an actuator cell on
the right.  These then diffuse their marker proteins \emph{CPT},
\emph{SPTO}, \emph{SPT1} and \emph{APT0} respectively.  In the
first time step, only the first gene (Fig.~\ref{sgenome}) is
active in the tissue seed and all the rest are suppressed. This
gene will always be active and step after step will split off
cells of type \emph{ACPT0} until all the surrounding hexagons are
occupied by these cells. After that, the seed does not execute any
further function other than secrete its own cell type protein
\emph{CPT}. The new cells will, in turn, also split off more cells
of type \emph{ACPT0} (gene 5), and so make the tissue grow larger
and larger (time=4 in Fig.~\ref{cellSto}). In a sense, these cells
provide a cellular support for further development of the actual
network, and could thus be called {\em glial}-type cells, in
analogy to the supportive function glial cells have in real
brains. These glial cells can split off three different types of
neurons. If the signal from the actuator \emph{APT0} is greater
than the signal from the tissue seed \emph{CPT}, then a neuron of
type \emph{ACPT3} will split off with probability $p>0$ (gene 2).
On the other hand, if the external protein signal of sensor
\emph{SPT0} is strong compared to the external protein of sensor
\emph{SPT1}, then instead a neuron of type \emph{ACPT1} will split
off with $p>0$ (gene 3). Last of all, if the signal \emph{SPT1} is
greater than \emph{SPT0}, then it is more likely that a neuron of
type \emph{ACPT2} will split off (gene 4). This is all that these
glial cells of type \emph{ACPT0} do: split off more glial cells,
or any of three differentiated neuron types depending on how close
they are to the sensors or the actuators.

These three cell types (\emph{ACPT1}, \emph{ACPT2} and
\emph{ACPT3}), will then form the actual neural network that will
do all the processing.  Through gene 6, cells of type \emph{ACPT1}
will grow a dendrite towards sensor \emph{SPT0} and define their
default neurotransmitter as \emph{NT1}.  In the same way, cells of
type \emph{ACPT2} will have gene 7 active and will grow a dendrite
towards sensor \emph{SPT1} and define their neurotransmitter as
\emph{NT2}.  Last of all, gene 8 is active in cells of type
\emph{ACTP3}, and will direct the growth of dendrites towards
cells of type \emph{ACPT1} and \emph{ACPT2} and an axon towards
the actuator \emph{APT0}.  In the end, each sensor \emph{SPT0} and
\emph{SPT1} is connected to every neuronal cell \emph{ACPT1} and
\emph{ACPT2}, and all the \emph{ACPT3} neuronal cells are
connected to the actuator \emph{APT0} (time=120 in
Fig.~\ref{cellSto}). However, which and how many \emph{ACPT1} and
\emph{ACPT2} neurons connect to which and how many of the
\emph{ACPT3} neurons relies on stochastic axonal growth,
preferably connecting neurons that are nearer on the hexagonal
grid. Moreover, all neurons end up connected after the axonal
growth process has finished, forming a fully functional NAND
implementation.

We still need to understand how the neurons actually process the
signals passing through them.  This is mediated through genes 9
and 10. Neurons \emph{ACPT1} and \emph{ACPT2} act as {\em relays}
of the sensor signals through gene 10. That is, whenever they
receive any neurotransmitter of type \emph{eNT} (default sensor
neurotransmitter) they will become excited and inject their
gene-defined neurotransmitters through their axons.  Neuronal
cells of type \emph{ACPT3} will then compute the NAND evaluative
action on the amount of neurotransmitters \emph{NT1} and
\emph{NT2} injected into their cell bodies and activate
accordingly (gene 9). Their activity causes the default
neurotransmitter to be injected into the actuator, thus finalizing
the simulated input-output NAND computation.

\section{Robustness of \emph{Stochastic}}

While \emph{Stochastic}'s neural tissue will always look different
every time it is grown because of the stochastic nature of
neuronal splitting, it always forms a processing network that
correctly computed the NAND function. This confers some robustness
to the phenotype of the network in spite of the stochastic, but
genetically directed, growth process.

\begin{figure}[h]
\begin{center} \setlength{\tabcolsep}{0.2mm}
\begin{tabular}{cc}
\includegraphics[width=1.65in]{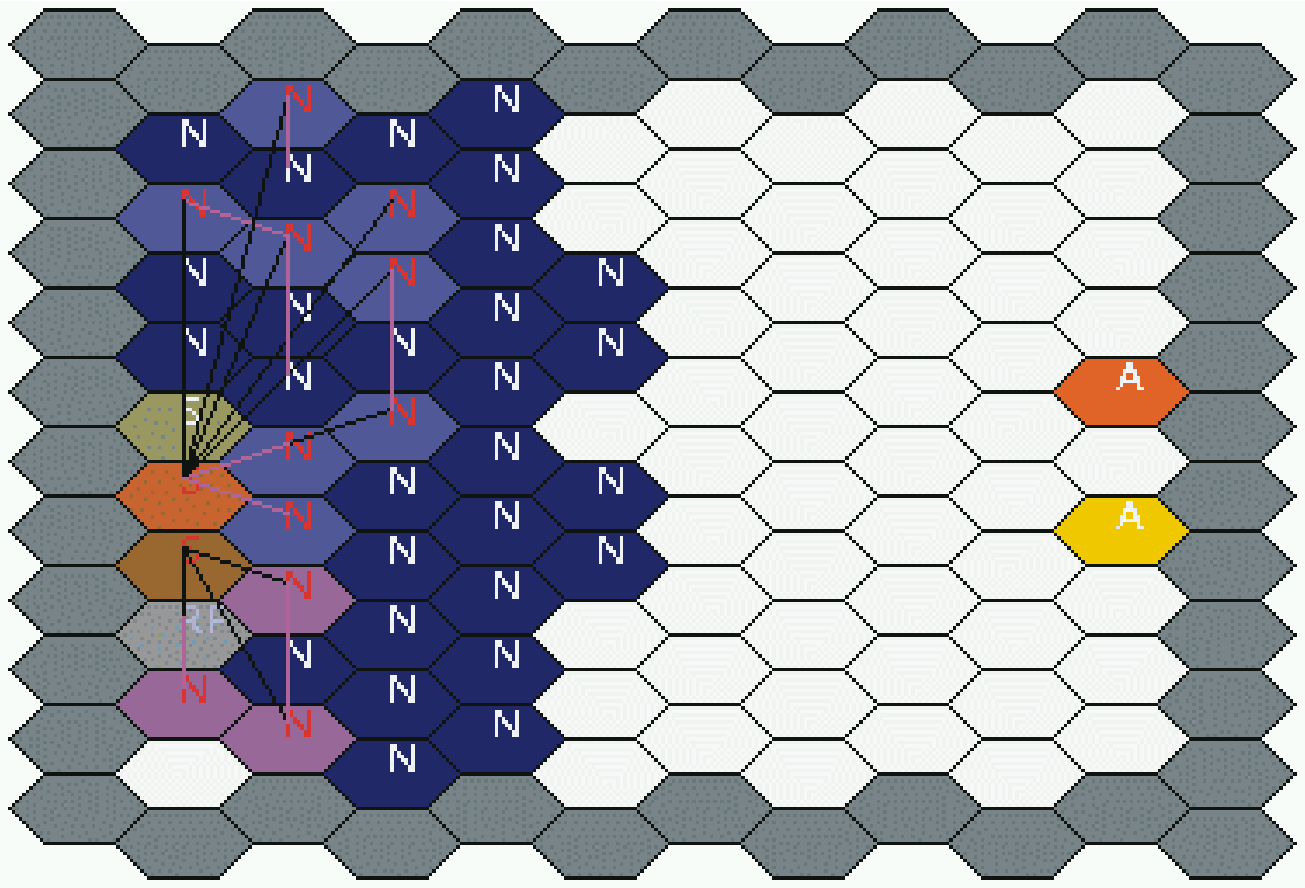} &
\includegraphics[width=1.65in]{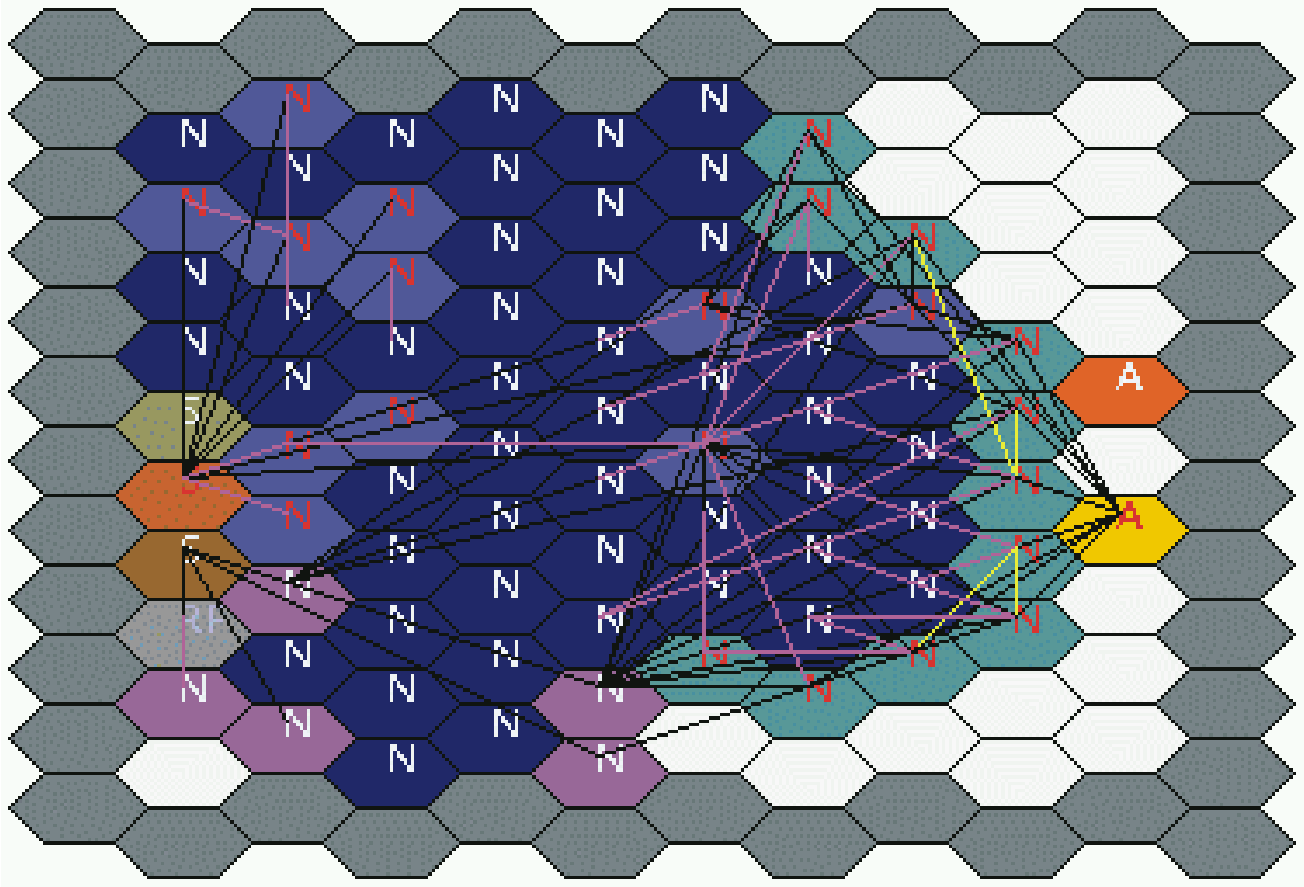}
\end{tabular}
\caption{Robustness of \emph{\textbf{Stochastic}} under cell
death. Half the neural tissue from Fig.~\ref{cellSto} was removed
(left). After $80$ time steps a different, but functional tissue
arises (right).} \label{RobustSto}
\end{center}
\end{figure}

However, the developmental process is far more robust than that.
For example, we can manually kill (remove) neurons of a fully
developed tissue and have a similar functional (but somewhat
scarred) tissue grow back. Fig~\ref{RobustSto} shows an example
where we even removed the tissue seed \emph{CPT}, which has an
important role in the organisms development (without its external
signal, glial cells of type \emph{ACPT0} do not proliferate).
While the morphology of the self-repaired tissue has changed, it
still computes the NAND function.  More than anything, this
observation helps illustrate the potential capabilities of
developmental processes in artificial chemistries to create robust
information processing neural tissues even under the breakdown of
part of their structure.  Note that the self-repair property of
\emph{Stochastic} was not evolved (or even hand-coded), but rather
emerged as a property of the developmental process. Naturally,
these robustness traits can be augmented and exploited under
suitable evolutionary pressures.

\section{Evolution of organisms in Norgev}

\begin{figure}[h]
\begin{center}
\includegraphics[width=2.4in]{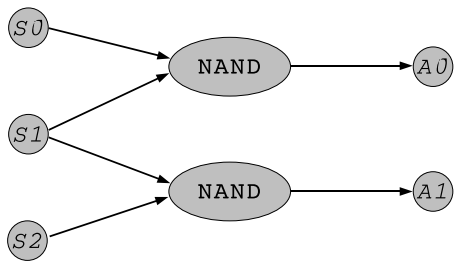}
\caption{Evolution objective: to \emph{double} the functionality
of the original organism.} \label{doubleNAND}
\end{center}
\end{figure}

Here, we present the evolutionary capabilities of Norgev, that is,
how its genetic structure and chemistry model allow for the
evolution of developmental neural networks that solve
pre-specified tasks. In the previous section we presented the
\emph{\textbf{Stochastic}} organism, which grew into a neural
tissue that computed a NAND function on its inputs. Our goal was
to study how difficult it would be to \emph{double} the tissue's
functionality and compute a double NAND on three inputs, and send
the result to two outputs (Fig.~\ref{doubleNAND}). Because one of
the mutational operators used in the Genetic Algorithm is {\em
gene doubling} (see Astor and Adami, 2000), we surmised that there
was an easy route through duplication and subsequent
differentiation. Because of the universality of NAND, showing that
more complex tissues can evolve from \emph{Stochastic} suggests
that arbitrary computational tissues can evolve in Norgev.

The input signal was applied for four time steps (the time for the
input to pass through the tissue and reach the output), and then
the output was evaluated by a reward function $R=1-\sqrt{\sum_{i}
(y_i(x) - t_i(x))^2}$ where $x$ is the input, $y$ the tissue's
output, and $t$ the expected output. Organisms were then selected
according to a fitness function given by the average reward over
400 time steps, and a small pressure for small genome sizes and
neuron numbers. Mutation rates were high and evolution was mainly
asexual. Details of the experiments will appear elsewhere (Hampton
and Adami, in preparation).

\begin{figure}[h]
\begin{center}
\includegraphics[width=3.25in]{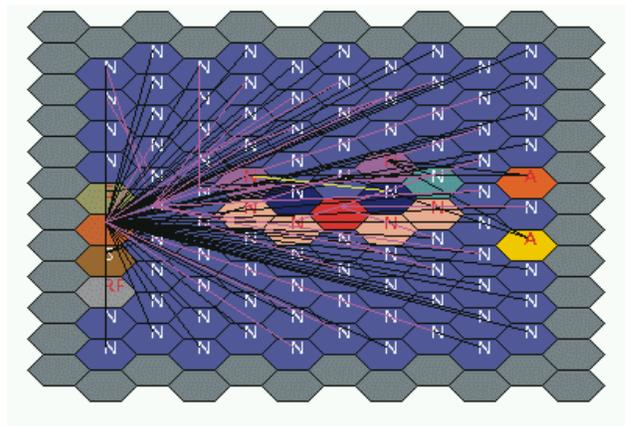}
\caption{\emph{\textbf{Stochastic A}} neural tissue expressing 6
different cell types. Most of the axonal connections that spread
out from the central sensor are not utilized. Instead, the actual
computation takes place in a compact area near the center.}
\label{cellStoA}
\end{center}
\end{figure}

We evolved organisms that obtained the double NAND functionality
in two separate runs on massively parallel cluster computers, over
several weeks. The two solutions were very different in both
structure and algorithm. The simplest, \textbf{\emph{Stochastic
A}}, evolved the fastest with the more straightforward morphology
(Fig.~\ref{cellStoA}).  Its genome is short (Fig.~\ref{sgenomeA})
when compared to evolved organisms in other runs, but is
substantially more difficult to understand compared to its
ancestor.

\begin{figure*}[tb]
\begin{center}
\colorbox[gray]{0.9}{\scriptsize
\begin{tabular}{llll}
1. & MUL(\emph{cpt}) NNY(\emph{apt2}) \textcolor [gray]{0.4}{SUB(\emph{ep2})} & $\Rightarrow$ & SPL(\emph{acpt0}) SPL(\emph{apt2}) \textcolor [gray]{0.4}{GRA(\emph{ep2})} \textcolor [gray]{0.4}{DFN(\emph{NT1})} \\
2. & SUP(\emph{acpt0}) \textcolor [gray]{0.4}{SUB(\emph{spt3})} \textcolor [gray]{0.4}{SUB(\emph{ep2})} & $\Rightarrow$ & SPL(\emph{acpt2}) SPL(\emph{acpt1}) INH1 \textcolor [gray]{0.4}{GRA(\emph{acpt5})} \textcolor [gray]{0.4}{MOD-(\emph{NT1})} SPL(\emph{acpt3}) \\
6. & SUP(\emph{acpt1}) ANY(\emph{spt0}) ADD(\emph{cpt}) MUL(\emph{NT1}) ADD(\emph{acpt2}) & $\Rightarrow$ & GDR(\emph{spt0}) SPL(\emph{acpt1}) \textcolor [gray]{0.4}{GRA(\emph{ep2})} \textcolor [gray]{0.4}{GRA(\emph{acpt5})} \textcolor [gray]{0.4}{MOD-(\emph{NT1})} \\
7. & SUP(\emph{acpt2}) NAND(\emph{spt1}) NSUP(\emph{spt1}) ADD(\emph{acpt0}) & $\Rightarrow$ & GDR(\emph{spt1})  DFN(\emph{NT2}) \textcolor [gray]{0.4}{GRA(\emph{ep2})} GRA(\emph{apt1}) \\
8. & SUP(\emph{acpt3}) ANY(\emph{apt0}) & $\Rightarrow$ & GDR(\emph{acpt1})  GDR(\emph{acpt2})  GRA(\emph{apt0}) \\
10. & NAND(\emph{eNT}) \textcolor [gray]{0.4}{OR(\emph{ep2})} & $\Rightarrow$ & EXT0 \textcolor [gray]{0.4}{PRD(\emph{ip0})} \\
11. & ANY(\emph{acpt3}) NSUP(\emph{acpt3}) MUL(\emph{acpt1})
\textcolor [gray]{0.4}{NAND(\emph{NT2})} AND(\emph{acpt0})
NNY(\emph{rfp}) & $\Rightarrow$ & DFN(eNT) \textcolor
[gray]{0.4}{INH1} \textcolor [gray]{0.4}{MOD-(\emph{NT1})}
GRA(\emph{apt0})
\end{tabular}
} \caption{Genome of evolved \emph{\textbf{Stochastic A}}
organism. Gene numbering is maintained from the ancestral genome,
and gene 11 is a new gene which was randomly created.  Gene atoms
in light gray appear to be useless and are considered ``junk".}
\label{sgenomeA}
\end{center}
\end{figure*}

After careful analysis of the genome, paired with an evaluation of
the physical connections present in the neural tissue, we came to
the conclusion that the organism had not reused {\em any} genomic
material to double the NAND function, but had instead {\em
completely} rewritten its code to implement a shorter and more
efficient algorithm when compared to the ancestor we wrote. Let us
embark once again in a quick step-by-step genome analysis. Gene 1
is active in the tissue seed, which then splits off a cell of type
\emph{ACPT0} and \emph{APT2}. After this, the gene is forever shut
off because of the repressive \emph{NNY(apt2)} condition. Cell
\emph{ACPT0} then splits off cells of type \emph{ACPT1},
\emph{ACPT2} and \emph{ACPT3} through gene 2. This gene is always
active, and thus \emph{ACPT0} cells are always in an inhibitive
activation state (due to action atom \emph{INH1}). Gene 6 makes
\emph{ACPT1} cells grow a dendrite to sensor \emph{SPT0} and have
same-type daughter cells.  These are the cells that cover the
whole substrate in Fig.~\ref{cellStoA}. Gene 7 causes \emph{ACPT2}
cells to grow a dendrite towards sensor \emph{SPT1}, an axon
towards actuator \emph{APT1} and define its neurotransmitter as
\emph{NT2}. Through gene 8, \emph{ACPT3} cells grow a dendrite to
sensor \emph{SPT2}, a dendrite to cells \emph{ACPT2}, and an axon
to actuator \emph{APT1}.  Gene 11 is the most cryptic.  This gene
is only active in the first $\sim$3 time steps of the organism's
life, and effectively makes cells of type \emph{ACTP0},
\emph{ACPT1} (only the ones in the center, not all the rest) and
\emph{ACPT2} grow an axon towards the actuator \emph{APT0}. Once
the tissue has developed, gene 10 is used by all cells for
processing sensory information (neurotransmitter \emph{eNT}), on
which it performs a NOT function.

\begin{figure}[h]
\begin{center}
\includegraphics[width=2.7in]{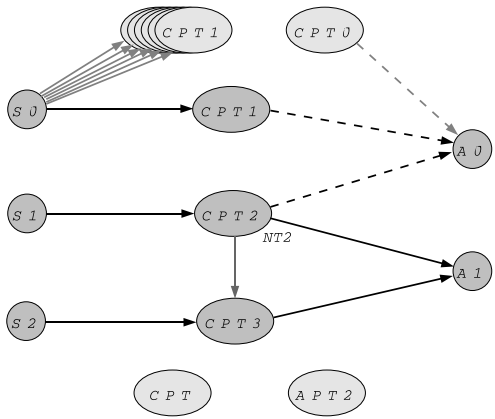}
\caption{Effective neural circuit grown by
\emph{\textbf{Stochastic A}}. Dashed axonal connections grow due
to gene 11, which is only active during the first moments of the
organisms life. Axons and neurons that have no influence on the
final computations are rendered in light gray.} \label{wireStoA}
\end{center}
\end{figure}

\begin{figure}[h]
\begin{center}
\includegraphics[width=2.4in]{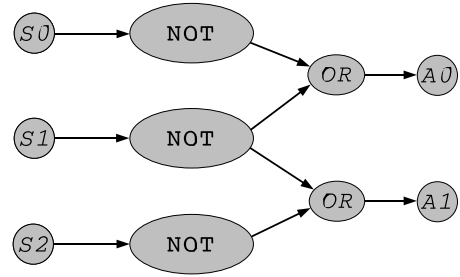}
\caption{The computation carried out by \emph{\textbf{Stochastic
A}.} \label{wireStoA_Logic}}
\end{center}
\end{figure}

The effective neural circuit is shown in Fig.~\ref{wireStoA}. The
result is processed in three time steps instead of the incorrectly
postulated minimum of four time steps. This is due to an implicit
OR function computed by the actuator cells that we did not
anticipate, but which was discovered and exploited by the
organism. The neural tissue is applying a NOT function at a relay
of its inputs, and then an OR on the actuators to arrive at the
double NAND (Fig.~\ref{wireStoA_Logic}). The resulting simplicity
of the organism is apparent from the fact that only gene 10 is
used for neural processing once the tissue has developed, and it
thus has a structure more conducive to further function doubling.

\begin{figure*}[t]
\begin{center}
\begin{tabular}{cccc}
$000 \rightarrow 11$ & $001 \rightarrow 11$ & $010 \rightarrow 11$
& $011 \rightarrow 10$ \\
\includegraphics[width=1.5in]{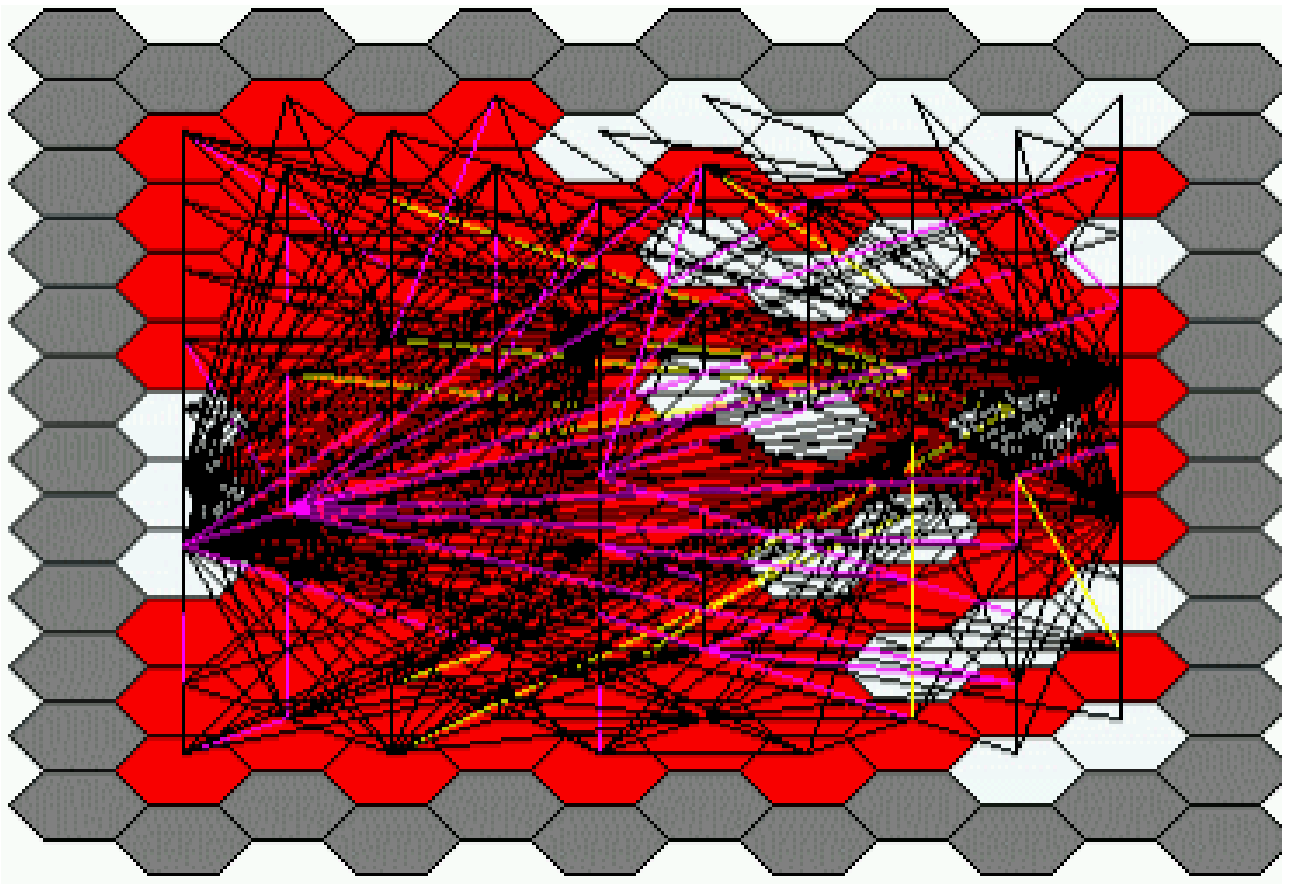} &
\includegraphics[width=1.5in]{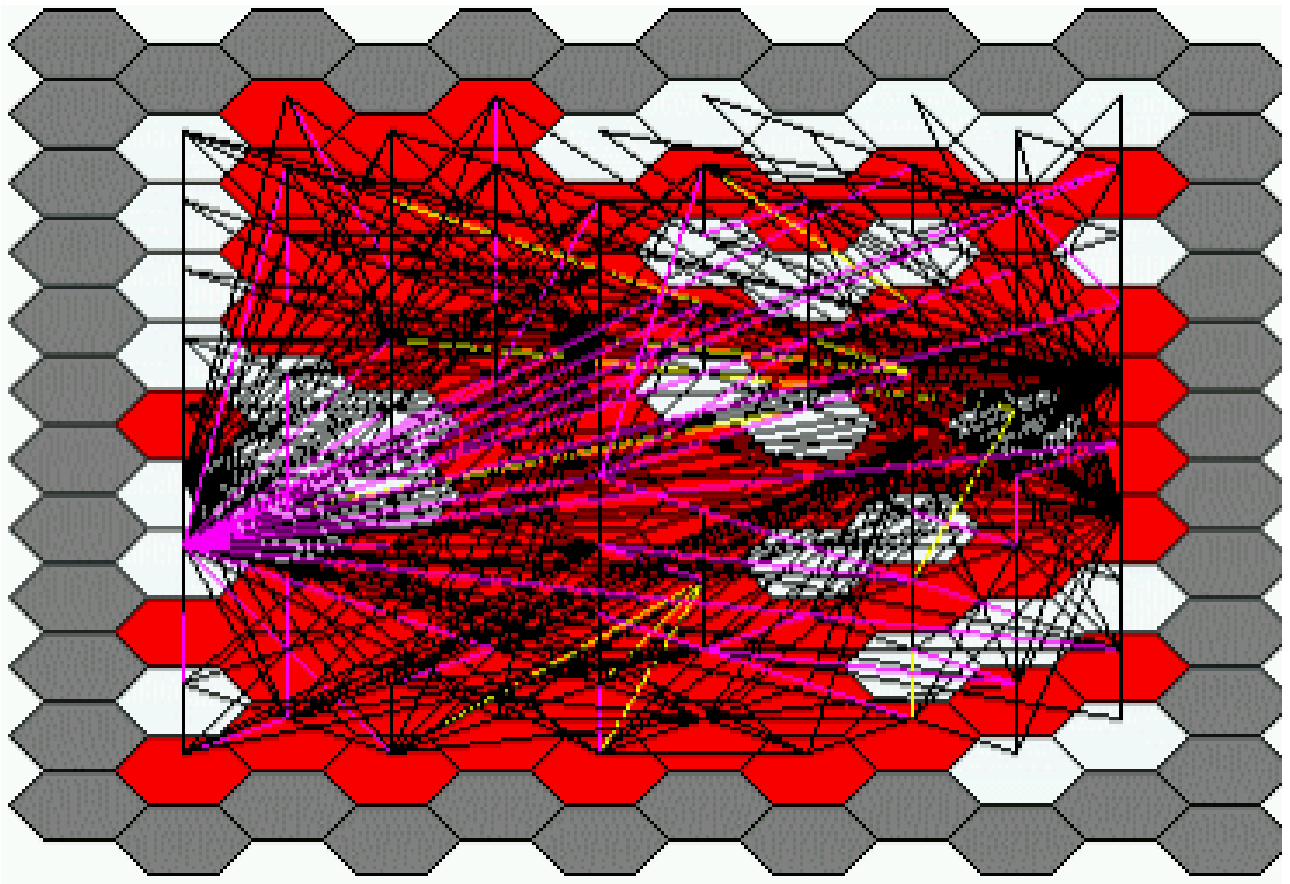} &
\includegraphics[width=1.5in]{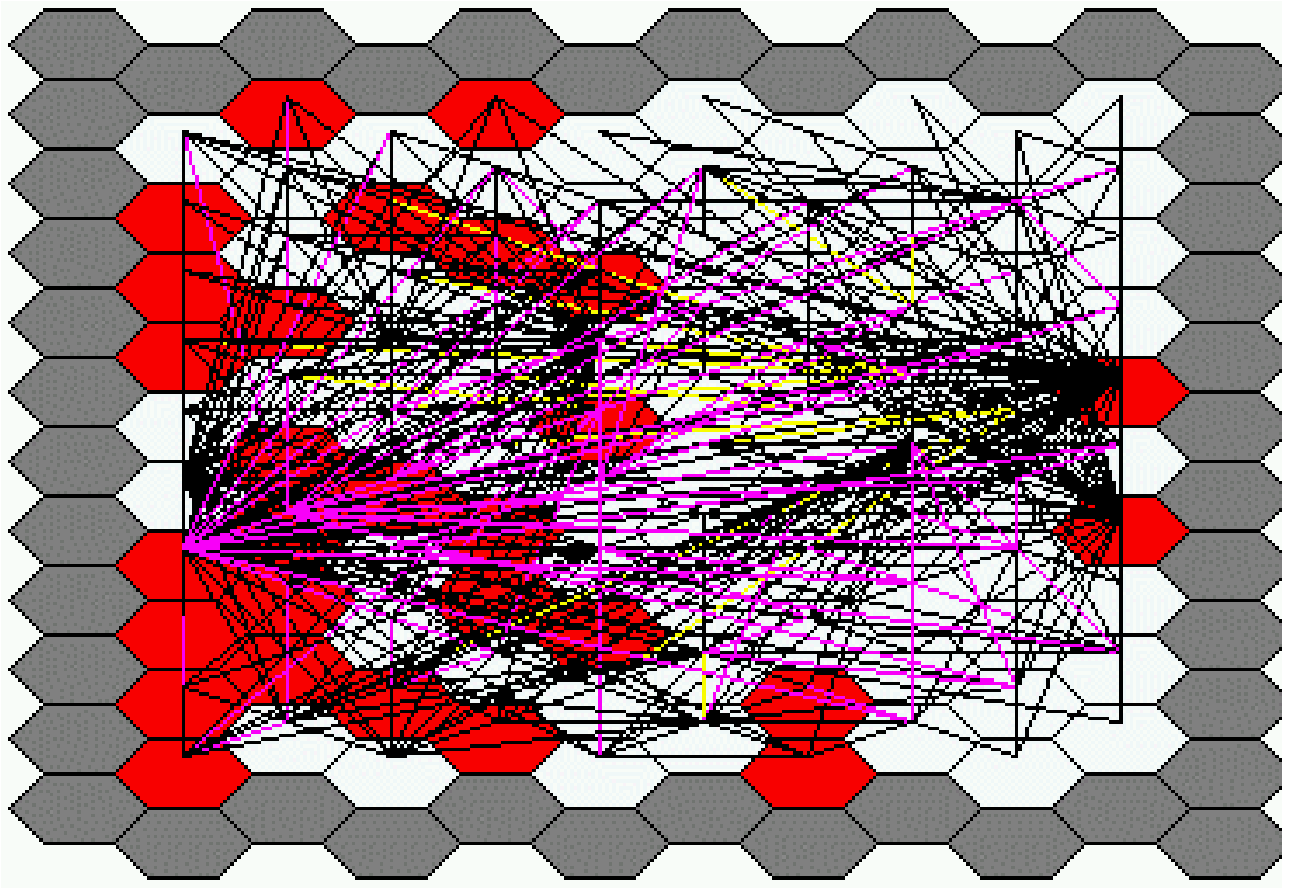} &
\includegraphics[width=1.5in]{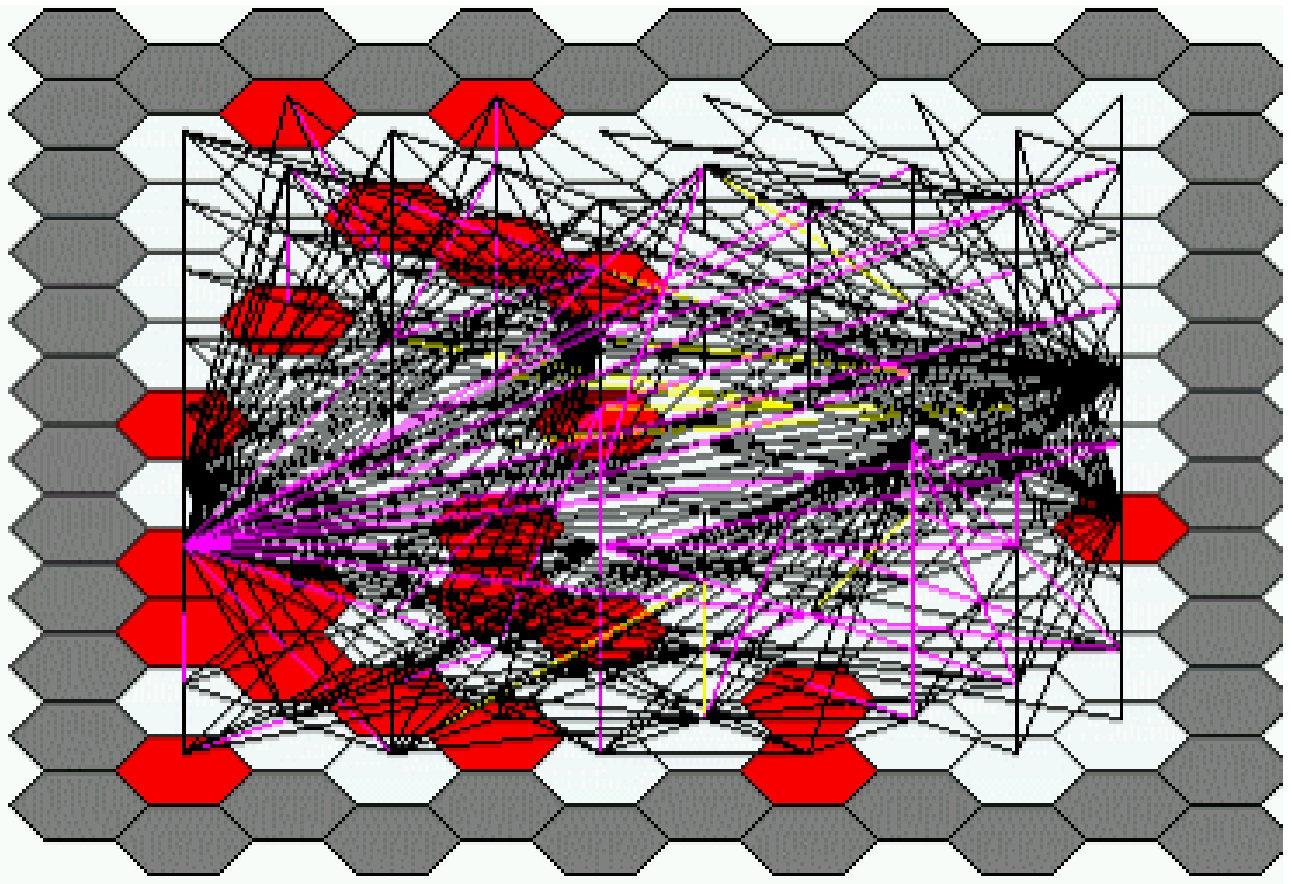} \\
$100 \rightarrow 11$ & $101 \rightarrow 11$ & $110 \rightarrow 01$
& $111 \rightarrow 00$ \\
\includegraphics[width=1.5in]{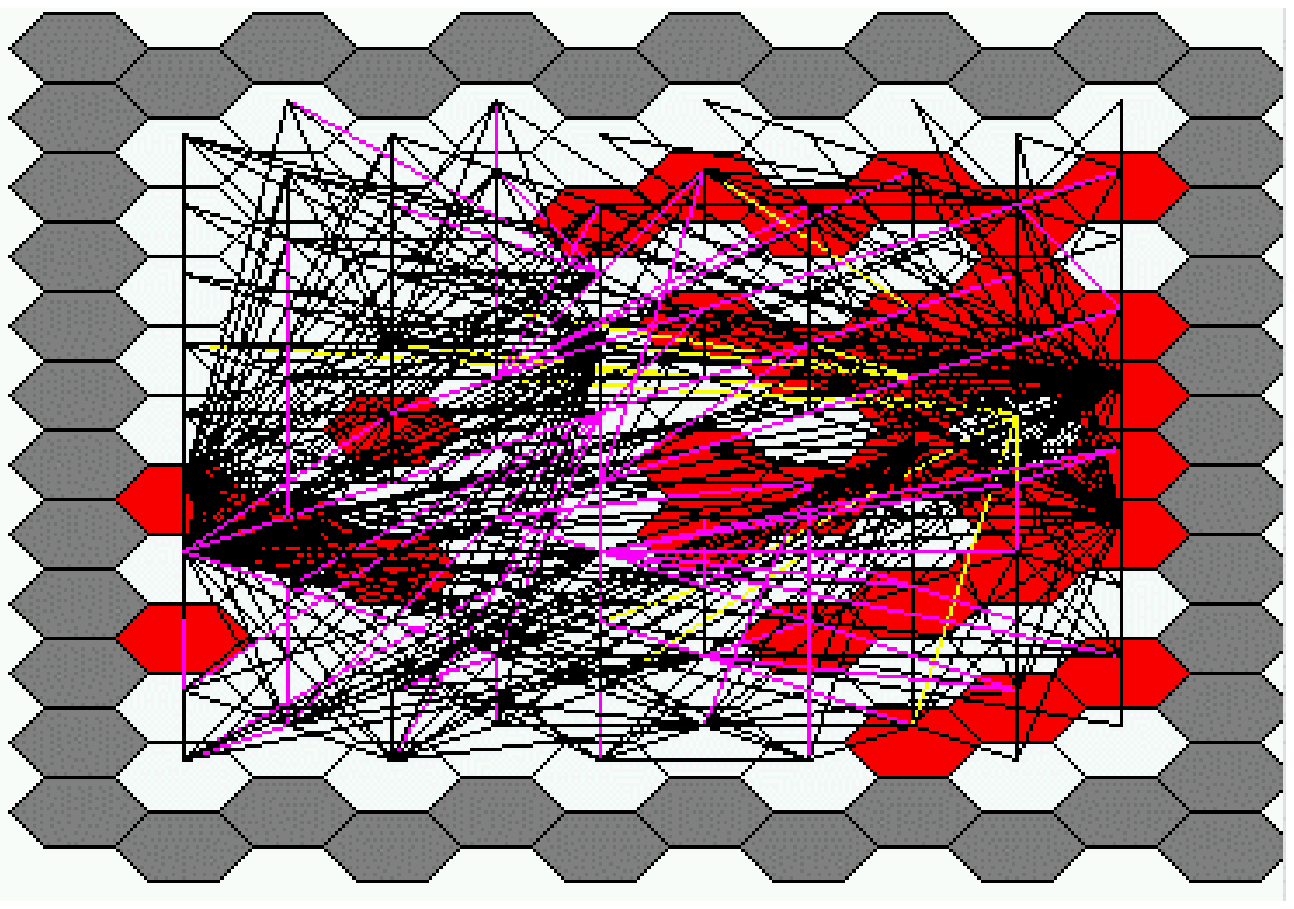} &
\includegraphics[width=1.5in]{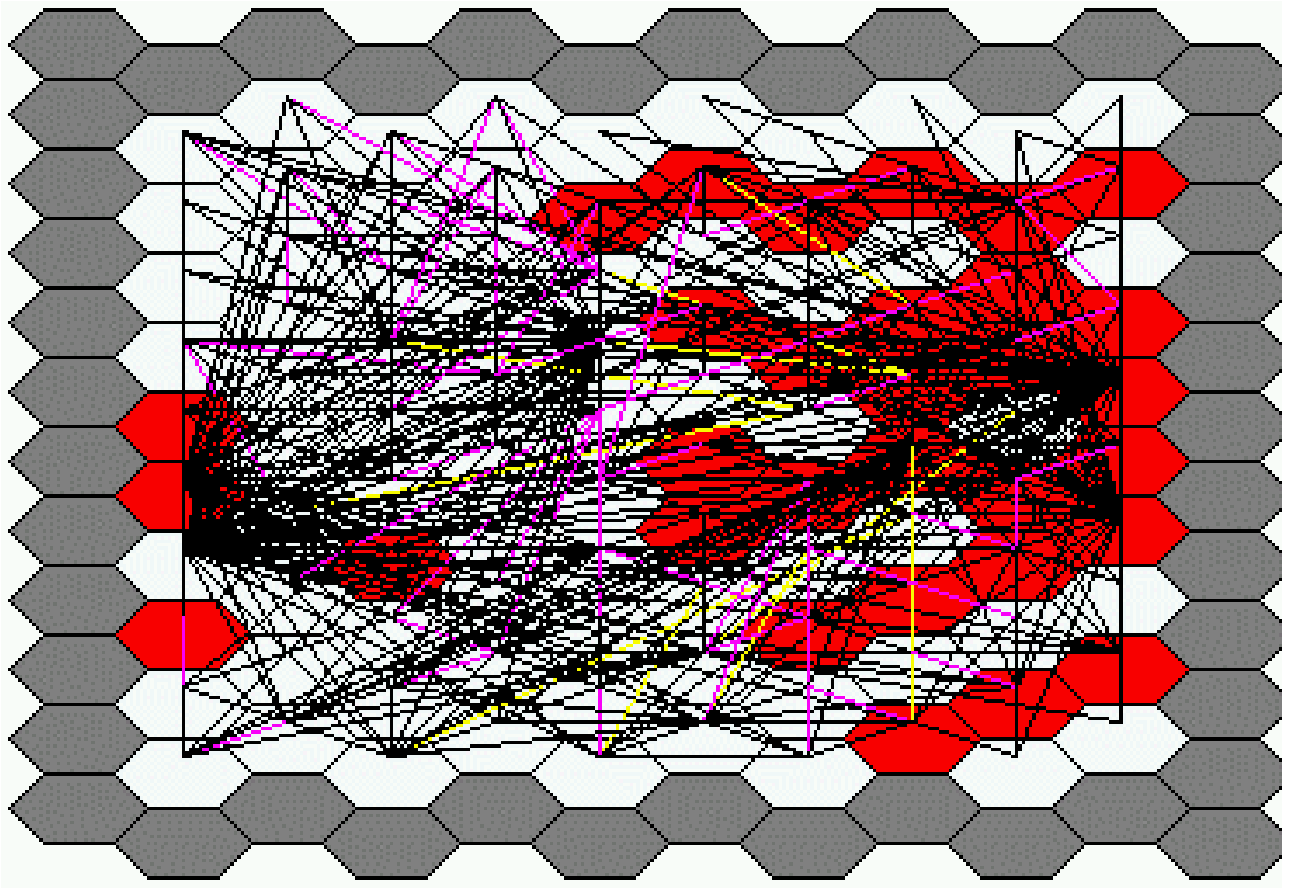} &
\includegraphics[width=1.5in]{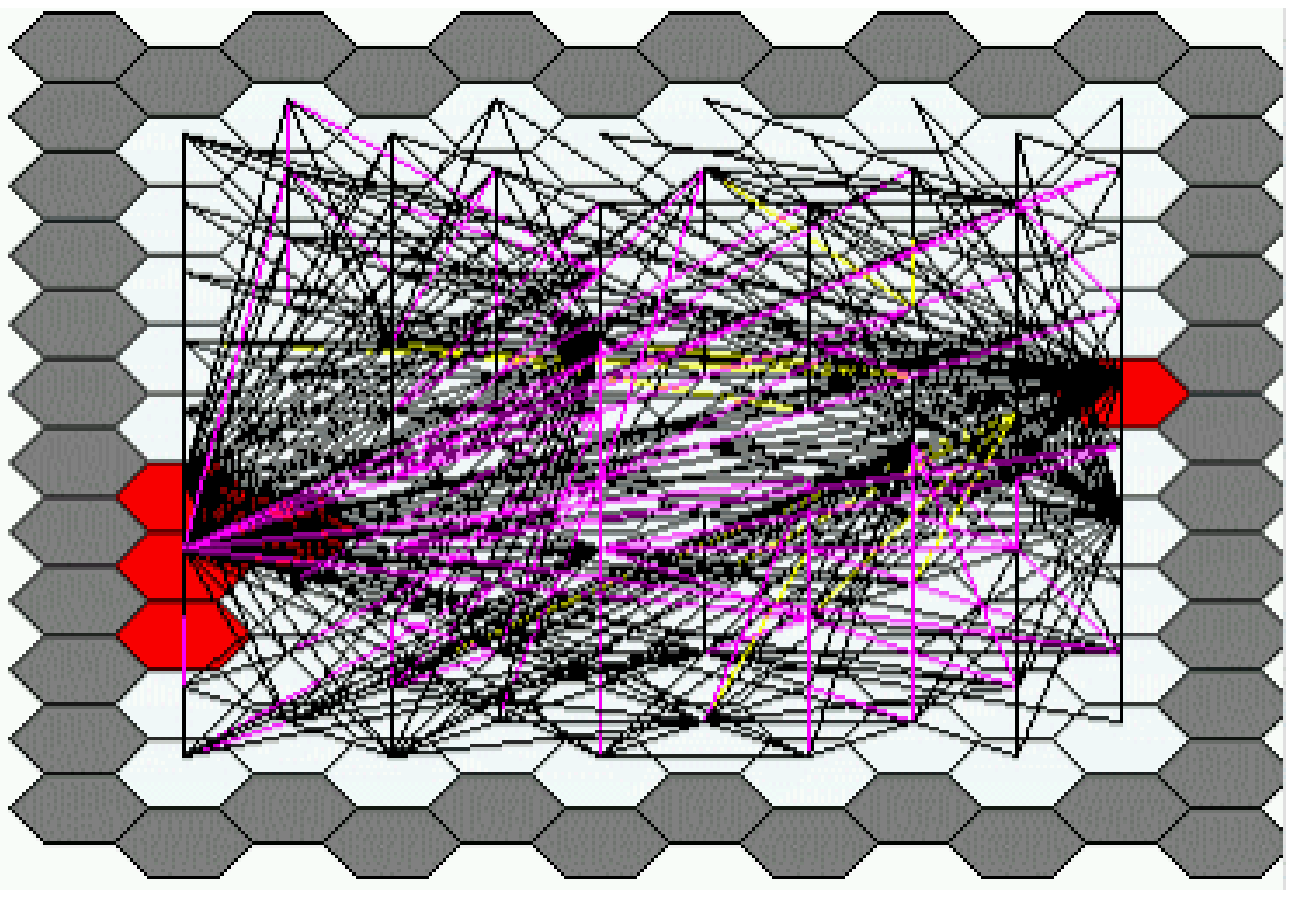} &
\includegraphics[width=1.5in]{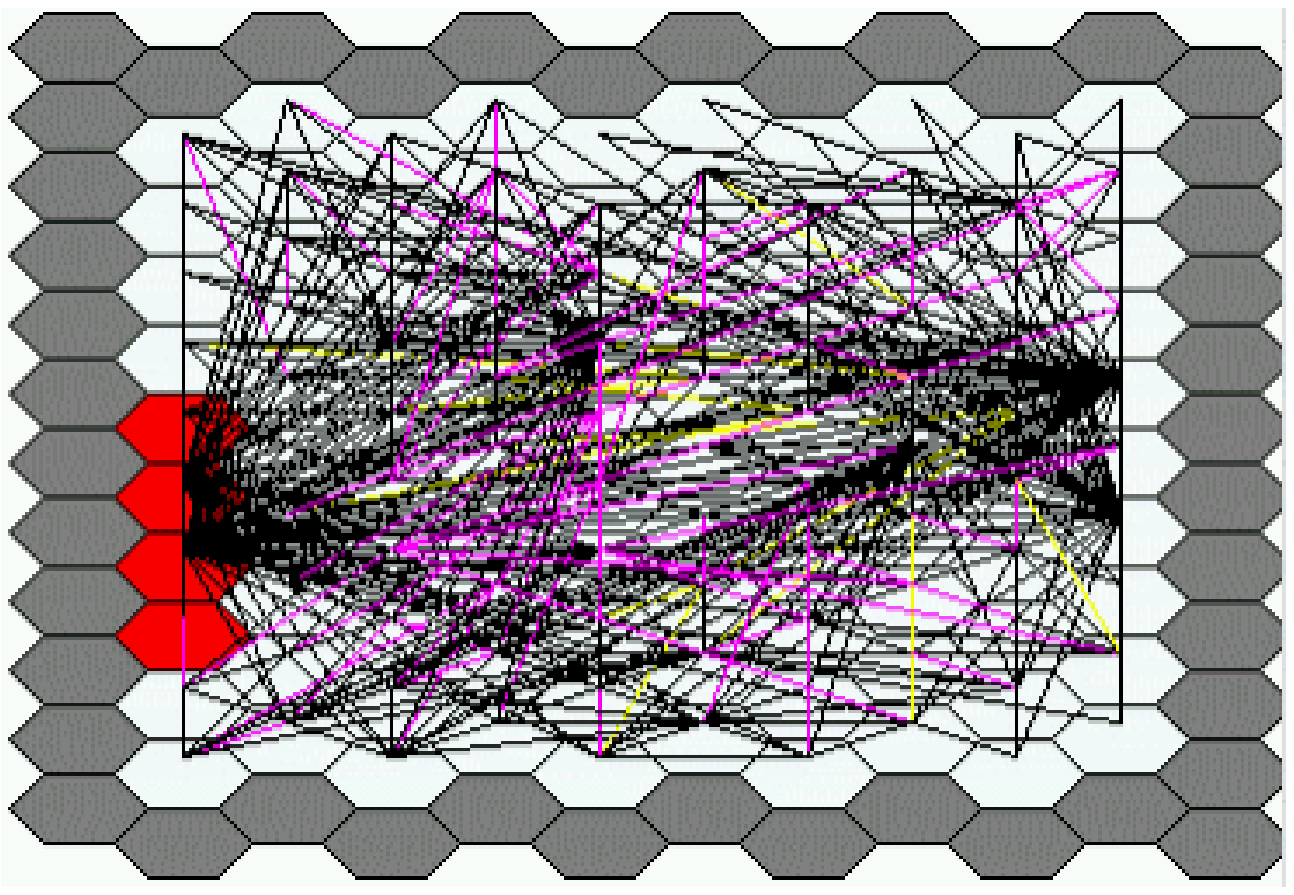} \\
\end{tabular}
\caption{Neuronal activity of a \emph{\textbf{Stochastic B}}
neural tissue under the eight possible binary input combinations,
where active neurons are shaded, and inactive neurons white. This
activity only reflects neurotransmitters that will be injected by
active neurons down their axonal branches.} \label{actvStoB}
\end{center}
\end{figure*}

Another organism that solved the problem was \textbf{\emph{
Stochastic B}}, which took considerably longer to evolve, and that
turned out to be highly complex and difficult to understand. In
Fig.~\ref{actvStoB}, cellular structures can clearly be seen in
which stripe-like patterns of two different neural types succeed
one another.  These stripes were different for each organism, and
reflect a stochastic development.  The axonal connections linking
all these neurons are so interwoven that it is difficult to
believe that this organism is actually acting on its inputs
instead of undergoing some recurrent neuronal oscillation.

We were unable to describe the development and internal workings
of this organism due to its complexity.  However, a complete
description is in principle always possible because of our access
to all of the organism's internal state variables, and more
importantly, to its genetic code: the source of its dynamics.
Taking the first steps in that direction, we studied the neuronal
activation under each of the eight possible input configurations
(Fig.~\ref{actvStoB}). We can clearly see neuronal activity that
follows the striped pattern on the right-hand side of the tissue
(for inputs of the form $x0x \rightarrow 11$). Remarkably, the
left side of the tissue does not follow the same organization and
thus we theorize that although they have the same cell type, they
have differentiated internally even further depending on their
position on the tissue. We came to the conclusion that this
organism is not performing the same internal computation as
\textbf{\emph{Stochastic A}}. We can see this by inspecting input
$110 \rightarrow 01$, and noticing that no tissue neurons are
activated, and thus there is no neuron performing the NOT function
on the last input.

\section{Conclusions}

Biology baffles us with the development of even seemingly simple
organisms. We have yet to recreate insect neural brains that
perform such feats as flight control. As an even simpler organism,
the flatworm {\it C. elegans}, has a nervous system which consists
of 302 neurons, highly interconnected in a specific (and mostly
known) pattern, and 52 glial cells,
but whose exact function we still do not understand. Within
Norgev, we have shown that such structural biocomplexity can arise
\emph{in silico}, with dendritic connection patterns surprisingly
similar to the seemingly random patterns seen in {\it C. elegans}.
And we might have been baffled at the mechanism of development and
function of our in silico neural tissue if it were not for our
ability to probe every single neuron, study every neurotransmitter
or developmental transcription factor, and isolate every part of
the system to understand its behavior. Thus, we believe that
evolving neural networks under a developmental paradigm is a
promising avenue for the creation and understanding of robust and
complex computational systems that, in the future, can serve as
the nervous systems of autonomous robots and rovers.

\section{Acknowledgements}

Part of this work was carried out at the Jet Propulsion
Laboratory, California Institute of Technology, supported by the
Physical Sciences Division of the National Aeronautics and Space
Administration's Office of Biological and Physical Research, and
by the National Science Foundation under grant DEB-9981397.
Evolution experiments were carried out on a OSX-based Apple
computer cluster at JPL.

\bibliographystyle{alife9}

\begin{thebibliography}{}

\bibitem[Astor and Adami, 2000]{Astor00}
Astor, J. and Adami, C. (2000).
\newblock A developmental model for the evolution of artificial neural
  networks.
\newblock {\em Artificial Life}, 6:189--218.

\bibitem[Belew, 1993]{Belew93}
Belew, R.~R. (1993).
\newblock Interposing an ontogenetic model between genetic algorithms and
  neural networks.
\newblock In {\em NIPS5 ed J Cowan (San Mateo), CA: Morgan Kaufmann}.

\bibitem[Dittrich et~al., 2001]{Dittrich01}
Dittrich, P., Ziegler, J., and Banzhaf, W. (2001).
\newblock Artificial chemistries--{A} review.
\newblock {\em Artificial Life}, 7:225--275.

\bibitem[Eggenberger, 1997]{Eggenberger97}
Eggenberger, P. (1997).
\newblock Creation of neural networks based on developmental and evolutionary
  principles.
\newblock In {\em Proc. ICANN'97, Lausanne, Switzerland, October 8-10, 1997}.

\bibitem[Gruau, 1995]{Gruau95}
Gruau, F. (1995).
\newblock Automatic definition of modular neural networks.
\newblock {\em Adaptive Behaviour}, 3:151--183.

\bibitem[Kirschner and Gerhart, 1998]{Gerhart98}
Kirschner, M. and Gerhart, J. (1998).
\newblock Evolvability.
\newblock {\em Proc. Natl. Acad. Sci. USA}, 95:8420--8427.

\bibitem[Kitano, 1990]{Kitano90}
Kitano, K. (1990).
\newblock Designing neural network using genetic algorithm with graph
  generation system.
\newblock {\em Complex Systems}, 4:461--476.

\bibitem[Koza et~al., 2003]{Koza03}
Koza, J.~R., Keane, M.~A., and Streeter, M.~J. (2003).
\newblock The importance of reuse and development in evolvable hardware.
\newblock In {\em 5th NASA/DoD Workshop on Evolvable Hardware, Chicago, IL,
  USA}. IEEE Computer Society.

\bibitem[Koza and Rice, 1991]{Koza91}
Koza, J.~R. and Rice, J.~P. (1991).
\newblock Genetic generation of both the weights and architecture for a neural
  network.
\newblock {\em IEEE Intl. Joint Conf. on Neural Networks}, 2:397--404.

\bibitem[Nolfi and Parisi, 1995]{Nolfi95}
Nolfi, S. and Parisi, D. (1995).
\newblock Evolving artificial neural networks that develop in time.
\newblock In {\em Advances in Artificial Life, Proceedings of the Third
  European Conference on Artificial Life}, pages 353--367. Springer.

\end{thebibliography}

\end{document}